\documentstyle[11pt,epsfig]{article}
\begin{document}
\newcommand{\roughly}[1]%
       
\newcommand{\PSbox}[3]{\mbox{\rule{0in}{#3}
\includegraphics{#1}\hspace{#2}}}
\newcommand\lsim{\roughly{<}}
\newcommand\gsim{\roughly{>}}
\newcommand\CL{{\cal L}}
\newcommand\CO{{\cal O}}
\newcommand\half{\frac{1}{2}}
\newcommand\beq{\begin{eqnarray}}
\newcommand\eeq{\end{eqnarray}}
\newcommand\eqn[1]{\label{eq:#1}}
\newcommand\intg{\int\,\sqrt{-g}\,}
\newcommand\eq[1]{Eq.~(\ref{eq:#1})}
\newcommand\meN[1]{\langle N \vert #1 \vert N \rangle}
\newcommand\meNi[1]{\langle N_i \vert #1 \vert N_i \rangle}
\newcommand\mep[1]{\langle p \vert #1 \vert p \rangle}
\newcommand\men[1]{\langle n \vert #1 \vert n \rangle}
\newcommand\mea[1]{\langle A \vert #1 \vert A \rangle}
\newcommand\bi{\begin{itemize}}
\newcommand\ei{\end{itemize}}
\newcommand\be{\begin{equation}}
\newcommand\ee{\end{equation}}
\newcommand\bea{\begin{eqnarray}}
\newcommand\eea{\end{eqnarray}}

\def\Dsl{\,\raise.15ex \hbox{/}\mkern-12.8mu D}
\newcommand\Tr{{\rm Tr\,}}
\thispagestyle{empty}
\begin{titlepage}
\begin{flushright}
CALT-68-2849\\
\end{flushright}
\vspace{1.0cm}
\begin{center}
{\LARGE \bf  Mathematical Constraints on Financially Viable Public Policy }\\ 
\bigskip
~\\
\bigskip
\bigskip
{Martin Gremm$^a$ and Mark B. Wise$^b$ } \\
~\\
\noindent
{\it\ignorespaces
(a) Pivot Point Advisors, LLC\\
 5959 West Loop South, Suite 333,
              Bellaire, TX 77401 \\


\bigskip   (b) California Institute of Technology, Pasadena CA 91125\\


}\bigskip
\end{center}
\vspace{1cm}
\begin{abstract}

Social Security and other public policies can be viewed as a series of cash in and outflows  that depend on parameters such as the age distribution of the population and the retirement age. Given forecasts of these parameters, policies can be designed to be financially stable, i.e., to terminate with a zero balance. If reality deviates from the forecasts, policies normally terminate with a surplus or a deficit. We derive constraints on the cashflows of robust policies that terminate with zero balance even in the presence of forecasting errors. Social Security and most similar policies are not robust. We show that non-trivial robust policies exist and provide a recipe for constructing robust extensions of non-robust policies. An example illustrates our results.  
\end{abstract}
\vfill
\end{titlepage}

\section{Introduction}
Social Security is running into financial difficulty, public and private pension funds are chronically underfunded, and many public policies cost the tax payer much more than anticipated. What is the reason for this unfortunate track record? Is it poor management? Is it faulty policy design? Or is it perhaps unavoidable that public policies turn into financial problems?

Entitlement programs and other public policies generally function by redistributing money. They collect taxes from one group and make payments to another in order to accomplish some social objective. A well-designed policy should cover its cost. The sum of all cash inflows should be sufficient to cover the sum of all payments so that the policy terminates with a zero balance rather than a deficit or surplus. We will refer to such policies as \emph{stable} policies.

To design a stable policy, it is necessary to forecast the parameters that determine the cashflows over the life of the policy. For Social Security these parameters include the average life expectancy, the demographic composition of the population, etc. With these forecasts in hand, the designer can set the level of taxation and benefit payments to ensure that the policy covers its cost.

The example of Social Security shows that a stable policy can still run into financial difficulties if reality differs from the forecasts used to design it. Actual cashflows depend on the parameters observed in the real world, not on the forecast values used to design the policy. This is the reason why Social Security is accumulating an ever-growing deficit. People live longer than forecast and the ratio of people paying in to people drawing benefits is lower than expected.

Even the best forecasts will differ from observed reality. Furthermore, the difference between the forecast and the observed reality generally increases with the age of the forecast. See Refs.~\cite{Keilman} and \cite{BA} for examples. Since such forecasting errors are unavoidable, a good policy should not only be stable, it should forgive forecasting errors. A \emph{robust} policy will still terminate with a zero balance even if reality differs from the forecast by a small amount\footnote{Our terminology is similar to the usage in Robust Control Theory. See, for example, \cite{robustcontrol}.}. 

In this paper we derive model-independent constraints on the cashflows of robust policies. We show that robust polices must have cashflows that depend not only on parameters but also on their rate of change. This allows such policies to adapt to small unanticipated changes in the parameters that determine the cashflows. We also briefly discuss a class of \emph{super robust} policies that can adapt to arbitrarily large forecasting errors.

The cashflows of most current policies depend only on parameters, but not on their rate of change. These policies cannot be robust. This is part of the reason why entitlement programs tend to run into financial difficulties. We provide a recipe for constructing robust extensions of such non-robust policies to show that such extensions always exist. 

In practice, it is very difficult to generate forecasts for the parameters that determine the policy cashflows. Ref.~\cite{kiato} presents a detailed model that relates the forecasts for some twenty-two parameters, ranging from tax rates to government debt and the value assigned to leisure time, to future Social Security contributions and payments. This model is very complex in part because it attempts to capture human behaviors such as the decision when to retire. These decisions in turn determine the cashflows into and out of the Social Security trust fund. Modeling human behavior requires modeling subjective judgments, for example the importance of leisure time. These quantities are difficult to measure directly. 

For our analysis, we assume that the cashflows are determined by directly measurable quantities and that forecasts for these quantities are available. For example, we would think of cash outflows from Social Security in terms of the number of retired people, a quantity can easily be measured and compared to a forecast. This model-independent approach allows us to avoid the complexities of generating forecasts and enables us to make general statements about the parameter dependence of policy cashflows.

In the next section we define the mathematical framework for this discussion. The following two sections discuss the general properties of robust policies and show that current policy design generally yields non-robust policies. Section \ref{sect_robustexamples} introduces simple examples of robust and super-robust policies. In Section \ref{robustextension} we show that it is always possible to find  a robust extension of a non-robust policy. Sections \ref{sect_toySS} and \ref{sect_robusttoyss} illustrate the results from earlier sections by applying them to a simple example.  We conclude in section \ref{sect_summary} with a summary and some remarks.

\section{Definitions}
\label{sect_def}

The net value of the policy at time $t$ is the sum of the cash in and outflows, $c_{i}$ and $c_{o}$, from inception of the policy at $t^\prime=0$ to $t^\prime=t$:

\begin{equation}
\label{cfdef}
V(t) = \int_{0}^t dt^{\prime} (c_{i}-c_{o}) e^{r(t-t^{\prime})}.
\end{equation}

For simplicity, we calculate the future value of the cashflows assuming that both surpluses and deficits grow at a fixed rate $r$. The definition of $V(t)$ ensures that $V(0)=0$, i.e., the policy begins without preexisting assets or liabilities. It then accumulates net assets or liabilities as time progresses.

The value of the policy depends on the difference between cash in and outflows. The constraints we derive in this paper only apply to this net cashflow, leaving it up to the policy designer to determine the split between cash in and outflows. In practical terms, a policy designer would probably start from the desired payout, $c_{o}$, to accomplish the policy goal, and construct a $c_{i}$ such that the robustness constraints on the net cashflow are satisfied. 

The cashflows $c_{i/o}$ are functions of $t^\prime$ and depend on one or more parameter functions. For entitlement programs these parameter functions could include the life expectancy, the percentage of people receiving benefits, and similar quantities.

In this framework, a \emph{stable} policy (as defined in the introduction) terminates at time $T$ with a zero balance, $V(T) = 0$, assuming that reality matches the forecast. A \emph{robust} policy terminates with $V(T)=0$ even if there are small differences between the observed values and the forecast. A \emph{super robust} policy terminates at $V(T)=0$ no matter how much reality differs from the forecast.

This discussion requires two versions of each of the parameter functions, the forecast and the observed values. We use $p_k$, where $k$ enumerates the parameter functions, for the forecasts that are available at inception of the policy. A policy designer determines the parameter dependence of the cashflows $c_{i/o}$ using these forecasts, because the observed values values are not yet known.

When the policy is implemented, the parameter dependence of the cashflows remains fixed, but their values are determined by the observed parameter functions, rather than the forecasts. We will use $\hat{p}_k$ for the observed values corresponding to the forecasts $p_k$. The observed values determine the value of the real-world policy cashflows.

The focus of this paper is to derive constraints on robust policy design. At design time, the policy designer only knows the forecasts and that reality will differ in some unknown way from  these forecasts. The next few sections discuss constraints that can be derived using this knowledge. 

Policy cashflows can depend on the $p_k$ at various points in time. For example, Social Security payouts may depend on demographic data (life expectancy, population size, etc) at some earlier time. Define a generalized parameter function to use in policy design as 

\begin{equation}
q_k(t^\prime) = \int_{-\infty}^{t^\prime} dt^{\prime\prime} F_k(p_k(t^{\prime\prime}),t^{\prime\prime} )g_k(t^\prime,t^{\prime\prime}),
\end{equation}
where $F_k$ is a functional of the parameter $p_k$ and $g_k$ is a weight function. 
The observed $\hat{q}_k$, calculated from the $\hat{p}_k$, determine the actual cashflows. The definition of the $\hat{q}_k$ ensures that they depend only on values of $\hat{p}_k$ that are known at time $t^\prime$.

Armed with these generalized parameter functions, we can now introduce a convenient notation for the adjusted net cashflows\footnote{A dot over a function of time represents its time derivative.} in Eq.~(\ref{cfdef}):

\begin{equation}
\label{defQ}
Q(q_k,\dot{q}_k, t^\prime)=(c_{i}-c_{o}) e^{r(T-t^{\prime})}.
\end{equation}

The cashflows are a function of the generalized parameter functions $q_k(t^\prime)$, their rate of change, and $t^\prime$. This defines a fairly general class of cashflows that includes many current policies as well as a wide range of non-trivial robust examples. 

We have made a two simplifying assumptions. $Q$ can depend on higher time derivatives of $q_k$. Including them would add flexibility for real policy design, but it would complicate the presentation here without adding anything qualitatively new.

A more important simplification is our assumption that there is a one-to-one correspondence between the observable parameters $p_k$ and the generalized parameters $q_k$.  Many real policies fall into this category, but more general cashflows can depend on different averages of the same parameter function. We do not consider this situation in this paper.

\section{From Stability to Robustness}
\label{sect_constraint}

Given $p_k$, a policy designer can construct $q_k$ and $Q$ so that the policy terminates with a net zero balance and accomplishes the policy goals. The resulting policy is stable by design. 

In reality, the policy cashflows depend on the observed values of the parameters, $\hat{p}_k$, which normally differ from the forecasts. Replacing the forecast with the observed values can be viewed as a variation of the parameter functions, $p_k$:

\begin{equation}
\label{variation}
p_k \rightarrow \hat{p}_k= p_{k}+\delta p_k.
\end{equation}

Here $\delta p_k$ is the forecasting error. The forecasting error is non-zero only for $t^{\prime\prime}>0$. For $t^{\prime\prime}\le 0$, the observed reality is known at design time, so $p_k=\hat{p}_k$. We assume the $\delta p_k$ are small and work to first order in these small quantities. 

Varying $p_k$ causes the generalized parameter functions to change as $q_{k}\rightarrow \hat{q}_k = q_k +\delta q_k$, where\footnote{We assume that the relation between $q_k$ and $p_k$ is such that there exist a $\delta p_k$ that can produce an arbitrary variation in $q_k$.}

\begin{equation}
\delta q_k = \int_{-\infty}^{t^{\prime}} dt^{\prime\prime}\frac{\partial F_k(p_{k}(t^{\prime\prime}),t^{\prime\prime} )}{\partial p_{k}} \delta p_k(t^{\prime\prime})g_k(t^\prime,t^{\prime\prime})+ O(\delta p^2).
\end{equation}

Similarly, for cashflows that depend only on $q_k$, $\dot{q}_k$ and $t'$, we find $Q\rightarrow  Q+\delta Q$ with 

\begin{equation}
\label{deltaQ}
\delta Q = \sum_k \left( \frac{\partial Q}{\partial q_{k}}\delta q_k+\frac{\partial Q}{\partial \dot{q}_{k}}\delta \dot{q}_k \right) + O(\delta q^2).
\end{equation}
This finally lets us compute the change in the terminal value of the policy as we replace the forecast parameter function with the observed one:

\begin{equation}
\delta V = \int_0^T dt^\prime \sum_k \left( \frac{\partial Q}{\partial q_{k}}\delta q_k+\frac{\partial Q}{\partial \dot{q}_{k}}\delta \dot{q}_k \right) + O(\delta q^2).
\end{equation}
Integrating by parts, we find

\begin{equation}
\delta V =\int_0^T dt^\prime \sum_k \left(  \frac{\partial Q}{\partial q_{k}}-\frac{d}{dt^\prime}\frac{\partial Q}{\partial \dot{q}_{k}} \right)\delta q_k + \frac{\partial Q}{\partial\dot{q_k}}\delta q_k {\Big |}_{t'=T} + O(\delta q^2),
\end{equation}
where the boundary term reflects that in general  the $\delta q_k (T)$ are not equal
to zero.

A robust policy must have $\delta V=0$, i.e., its terminal value must remain unchanged if reality differs from the forecast. Policies are robust (at leading order) if the net cashflows satisfy the following Euler-Lagrange equation\footnote{See, for example, \cite{gold}.} and boundary condition:

\begin{equation}
\label{constraint}
\left( \frac{\partial Q}{\partial q_k}-\frac{d}{dt^\prime} \frac{\partial Q}{\partial \dot{q}_k} \right)=0  \mbox{ and } \frac{\partial Q}{\partial\dot{q}_k}{\Big |}_{t^\prime=T} =0  .
\end{equation}

For this derivation we did not need to know the values of $\delta p$ or $\delta q$. We simply assumed that we have a non-zero forecasting error and that it is small enough to neglect higher order terms. Eq.(\ref{constraint}) provides a constraint on the functional form of the net cashflows $Q$ in terms of the forecasts $q_k$ available at design time. If the constraint is satisfied, the resulting policy will terminate with balanced books in the presence of small forecasting errors. 

\section{Policies Without Time Derivatives Cannot Be Robust}
\label{nonexistence}

The somewhat abstract constraint in Eq.~(\ref{constraint}) has far-reaching consequences for policy design. In this section we show that a broad class of policies that includes Social Security cannot be robust. 

The cashflows into and out of the Social Security fund depend only on the generalized parameter functions $q_k$, but not on their rate of change. For example, the inflows depend on the number of people making payments and how much they earn. They do not depend on how these quantities have changed. Similarly, the payouts depend on the number of recipients and how much they paid in while they were working. Again, the rate of change of these quantities has no bearing on the payments. 

Most current policies follow the same pattern. The cashflows depend on the generalized parameters $q_k$, but not on their time derivatives $\dot{q}_k$. For this type of policy $Q=Q(q_k,t^\prime)$. Evaluating the constraint, Eq.~(\ref{constraint}), for cashflows with this functional form we find

\begin{equation}
\label{nonRobustness}
\frac{\partial Q}{\partial q_k}=0.
\end{equation}

Non-trivial $Q$ that satisfy this constraint exist. For these cashflows, the leading order contributions to $\delta Q$ and $\delta V$ are of order $\delta q^2$ or higher.  Such policies achieve robustness by being approximately independent of $q_k$. Cashflows of the form $Q=f(t^\prime)$ are exactly independent of $q_k$. The special case $Q=0$, defines a Pay-As-You-Go policy for which the in and outflows net to zero at every point in time.

We conclude that there are no robust policies of the form  $Q=Q(q_k,t^\prime)$ with non-trivial $q_k$-dependance at leading order.

Most real-world policies depend on parameters. For example, a useful unemployment insurance should build up assets during times of full employment and pay them out when unemployment is high. To accomplish this, the net cashflow must be a function of the unemployment rate, i.e. $\partial Q/ \partial q \neq 0$, violating the constraint Eq.~(\ref{nonRobustness}). Social Security, defined benefit pension plans, and similar policies also have net cashflows that depend on parameters and therefore violate  Eq.~(\ref{nonRobustness}). Because the constraint is not satisfied, these policies cannot be robust. 

A policy that is exactly or approximately independent of the unemployment rate often fails to accomplish the policy goals. For example, a Pay-As-You-Go version of the unemployment insurance would either raise rates for people with jobs when unemployment is high to to cover payments to the unemployed or lower benefits to match the reduced inflows. A more general policy with $Q=f(t^\prime)$ would accumulate assets at a predefined time and have net outflows at other predefined times. Neither solution accomplishes the policy goal of mitigating the effect of economic cycles on the workforce.

In the introduction we asked why so many policies run into financial difficulties. Now we see that many policies are intrinsically unstable with respect to forecasting errors. This is a general result that only depends on the functional form of the net cashflows, not on any policy specific details. In order to construct non-trivial robust policies, the net cashflows must depend on time derivatives of the parameter functions.

\section{Non-Trivial Robust Policies}
\label{sect_robustexamples}

In the previous section we saw that it is impossible to construct non-trivial robust policies with cashflows of the form $Q(q_k,t^\prime)$. The cashflows had to be at least approximately independent of $q_k$ to satisfy the robustness constraint. In this section we show how to construct robust policies with non-trivial $q_k$ dependence. 

The easiest way to satisfy  Eq.~(\ref{constraint}) is to define $Q$ as a total time derivative of another function. A simple choice, $L(q_k,t^{\prime})$, yields

\begin{equation}
Q(q_k,\dot{q_k},t^{\prime}) = {d L \over dt^\prime}=\frac{\partial L}{\partial t^{\prime}} +\sum_k \frac{\partial L}{\partial q_k}\dot{q_k},
\end{equation}
with the boundary condition

\begin{equation}
\label{boundaryconditionL}
\frac{\partial L}{\partial q_k}{\Big |}_{t^{\prime}=T}=0.
\end{equation}

Any non-trivial $L$ that satisfies this boundary condition yields a robust policy with cashflows that depend on parameter functions. 

Policies with cashflows that are given by total time derivatives are especially tractable, because we can calculate the terminal value of the policy without knowing the functional form of $L$:

\begin{equation}
V(T) = \int_0^T dt^\prime Q =  L(q_k, t^\prime)  {\Big{ |}}_0^T.
\end{equation}

The boundary condition Eq.~(\ref{boundaryconditionL}) ensures that the variation of $V$ vanishes at linear order in $\delta q$:

\begin{equation}
\delta V = \sum_k \frac{\partial L}{\partial q_k}\delta q_k{\Big |}_{t^{\prime}=T} =0.
\end{equation}

By tightening the boundary condition on $L$ to require all derivatives to vanish at $t^\prime = T$, 

\begin{equation}
\label{superrobustbc}
\frac{\partial^n L}{\partial q_k^n}{\Big |}_{t^{\prime}=T}=0,
\end{equation}
we can identify a class of policies for which $\delta V$ vanishes to all orders in $\delta q$. These policies are guaranteed to terminate with $V(T)=0$ no matter how much reality differs from the forecast. They are super-robust.

It is straightforward to construct examples of super-robust policies. The ansatz below defines the cashflows in terms of an arbitrary function $M$ in a way that ensures that the condition for super-robustness, Eq.~(\ref{superrobustbc}), is satisfied.
\begin{equation}
Q(q_k, \dot{q_k},t')={1 \over T}{ d \over dt'} \left[ (T-t')M(q_k(t'),t') \right] +C(t')
\end{equation}
With this ansatz, the terminal value of the policy is given by
\begin{equation}
V(T)=-M(q_k(0),0)+\int_0^Tdt' C(t').
\end{equation}
$V(T)$ only depends on the value of $q_k$ at inception, which is known when the policy is designed. $C(t^{\prime})$ can be chosen to make $V(T)=0$. 

The results in this section show that it is mathematically very simple to construct robust and super-robust policies. However, a purely mathematical solution is likely to be of limited use for real-world applications that normally start from a policy goal rather than a desire to satisfy Eq.~(\ref{constraint}). The next section discusses ways to design robust policies based on more practical starting points.

\section{Robust Extensions of Non-Robust Policies}
\label{robustextension}

In Section \ref{nonexistence} we showed that most current policies are not robust, which contributes to the solvency problems many of them encounter. This section provides a recipe for constructing a robust extension of a non-robust policy. For simplicity we consider policies that depend on only one parameter.

We take the simplest approach possible to show that a robust extension always exists. This is one of infinitely many robust extensions to a given non-robust starting point. For real-world applications, the policy designer needs to find an extension that keeps the original policy goal intact without placing an undue burden on the population that is paying for it. 

Many existing policies are defined in terms of cashflows that depend on just the parameter function $q$, but not its time derivative. We can extend a non-robust policy $\tilde{Q}(q,t^\prime)$ by adding a correction term that allows us to satisfy the constraint Eq.~(\ref{constraint}). The simplest expression that leads to a robust policy is 

\begin{equation}
Q(q,\dot{q}, t^\prime)= \tilde{Q}(q,t^\prime)+A(t^\prime)\dot{q}+C(t^\prime).
\end{equation} 

Inserting this ansatz into the Euler-Lagrange equation, Eq.~(\ref{constraint}), the boundary condition becomes $A(T)=0$ and we find

\begin{equation}
\label{defA}
\frac{\partial\tilde{Q}}{\partial q}=\frac{dA}{dt^\prime}.
\end{equation}
This equation determines the time dependence of $A$. We emphasize that $A$ is a function of time only. It is set once at design time using the explicit time dependence of the forecast $q$ and does not change over the life of the policy. $C$ can be  chosen arbitrarily subject to the constraint that our modified policy should terminate with $V(T)=0$.

For a given $\tilde{Q}$ these equations always produce a robust policy. This robust policy is one of infinitely many ways to extend the non-robust starting point. Practical applications would most likely require a more flexible ansatz.

\section{Toy Social Security Policy}
\label{sect_toySS}

A very simplified ({\it i.e.}, toy)  Social Security policy illustrates the concepts discussed in the preceding sections. In this toy model, retirees receive a fixed payment,  $c_{out}$, and everybody else makes a fixed contribution,  $c_{in}$. $p(t^\prime)$ is the forecast  for the percentage of the population receiving benefits. We also assume that the total population is constant. Setting $q(t^\prime)=p(t^\prime)$, this gives us the following forecast value of the policy:

\begin{equation}
V(t)= \int_{0}^t dt^{\prime} (c_{in}(1-p(t^\prime))-c_{out}p(t^\prime)) e^  {r(t-t^{\prime})}.
\end{equation}
This is the Pay-As-You-Go solution where inflows equal outflows and the policy satisfies $Q=0$ for all $t^\prime$, if

\begin{equation}
\label{cio}
p(t^\prime)=\frac{c_{in}}{c_{in}+c_{out}}.
\end{equation}
This makes $p$ a constant, independent of time.

The observed percentage of retired persons, $\hat{p}=p+\delta p$, differs from the forecast by a forecasting error $\delta p$. To make the example explicit, we assume that $\delta p = \epsilon t^\prime$, where $\epsilon$ is a positive constant indicating that there is a higher percentage of retired people than expected at design time.

The actual value of the policy depends on the observed percentage of people in retirement:

\begin{equation}
V(t)= \int_{0}^t dt^{\prime} (c_{in}(1-\hat{p}(t^\prime))-c_{out}\hat{p}(t^\prime)) e^ {r(t-t^{\prime})}.
\end{equation}
Using our forecast, Eq.~(\ref{cio}), and the explicit form of the forecasting error we find

\begin{equation}
V(t)= -\epsilon \frac{c_{in}+c_{out}}{r^2}\left[ e^{rt}-1-rt\right] = -{\epsilon (c_{in}+c_{out})\over 2} t^2 +O(r),
\end{equation}
which  is negative and increases in magnitude with time. 

According to our definition of robustness, $V$ must remain unchanged to linear order in the presence of forecasting errors. The result above is linear in the forecasting error. Clearly this toy model is not robust.

\section{Rendering the Toy Social Security Policy Robust}
\label{sect_robusttoyss}

Consider the toy Social Security policy introduced in the previous section  with a finite  termination time, $T$,
\begin{equation}
{\tilde Q}=\left(c_{in}(1-p(t^\prime))-c_{out}p(t^\prime)\right)e^{r(T-t^\prime)}.
\end{equation}
As we saw earlier, this  policy is stable by design provided that $p$ satisfies Eq.~(\ref{cio}). In the previous section we also showed explicitly that the policy is not robust. In this section we show how the prescription in Section \ref{robustextension} allows us to find a robust extension of this toy model. 

Using that $q=p$ for this model, we find from Eq.~(\ref{defA}) that $A$ is given by:
 
 \begin{equation}
 {d A \over d t^\prime}=-(c_{in}+c_{out})e^{r(T-t^\prime)}.
 \end{equation}
 The solution that satisfies the boundary condition $A(T)=0$ is
  \begin{equation}
 A(t^\prime)=(c_{in}+c_{out})\left( {e^{r(T-t^\prime)}-1\over r}\right).
 \end{equation}
With these results we find the  following robust extension of our original policy:
\begin{equation}
Q=\left[ c_{in}(1-p(t^\prime))-c_{out}p(t^\prime)+(c_{in}+c_{out})\left({1-e^{-r(T-t^\prime)} \over r}\right)\dot{p}(t^\prime)\right]e^{r(T-t^\prime)}
\end{equation}
To further simplify the discussion we take the limit $r \rightarrow 0$:
\begin{equation}
\label{fixed}
Q=\left[ c_{in}(1-p(t^\prime))-c_{out}p(t^\prime)+(c_{in}+c_{out})(T-t^\prime)\dot{p}(t^\prime)\right].
\end{equation}

Now we repeat the calculation of the real-world value of the policy from the previous section to see how the robust extension responds to forecasting errors. As before, the observed percentage of retirees is $\hat{p}=p+\delta p$.
For the policy in Eq.~(\ref{fixed}) we then find that,
\begin{equation}
Q=(c_{in}+c_{out})(-\delta p(t^\prime)+(T-t^\prime)\delta\dot{ p}(t^\prime))=(c_{in}+c_{out}){d ((T-t^\prime)\delta p(t^\prime) )\over dt^\prime}.
 \end{equation}
 
This real-world cashflow integrates to $V(T)=0$ since $\delta p(0)=0$. This extension of our non-robust toy Social Security is manifestly robust. In fact is is super-robust because $Q$ is linear in $p$ in this example. More realistic, non-linear models will not have this property.

Of course there are other ways to make a policy robust, for example by simply modifying the inflows to match the outflows. This yields the Pay-As-You-Go solution with $Q=0$ discussed in Section \ref{nonexistence}.

It is informative to compare these two ways of constructing robust extensions of a non-robust starting point. Parametrize the robust extension as follows:

\begin{equation}
Q=D(p,\dot{p},t^\prime) (1-p(t^\prime))-c_{out}p(t^\prime),
\end{equation}
where the second term is the payout term from the original policy to preserve the policy goal, and the first term is a modified pay-in term to make the policy robust. $D$ is the amount working people have to pay in. It replaces the constant contribution $c_{in}$ in the non-robust starting point.

Using either the $Q$ from Eq.~\ref{fixed} or $Q=0$ for the Pay-As-You-Go solution, we can now solve for $D$ to find to leading order in $\delta p$

\begin{equation}
D_{PG} = c_{in}+\frac{(c_{in}+c_{out})^2}{c_{out}} \delta p
\end{equation}
in the Pay-As-You-Go case and 

\begin{equation}
D_{R}=c_{in}+\frac{(c_{in}+c_{out})^2}{c_{out}}(T-t^\prime)\delta \dot{p}
\end{equation}
for the robust extension in  Eq.~(\ref{fixed}).

$D_{PG}$ shows the disadvantages of the  Pay-As-You-Go solution. If the number of retirees exceeds forecasts, i.e., $\delta p>0$, the working population pays more. Conversely, if it is lower, they pay less. Pay-As-You-Go does not anticipate changes. It simply passes on the current cost of supporting the retirees to the working population. 

The detailed  properties of $D_{R}$ depend on how the robust extension is constructed, but even in our simple example, we can see how this robust policy anticipates change. At any point in time, the inflows are calculated based on how quickly reality diverges from the forecast, $\delta \dot{p}$, and the remaining time to termination, $T-t^\prime$. The policy extrapolates the current rate of divergence to the termination date, and adjusts the cash inflows based on this extrapolation. 

Both robust extensions have advantages and disadvantages. The Pay-As-You-Go solution tends to be cyclical, because it raises rates when the situation becomes difficult. If the forecasting errors are small, this is a good solution, but if they are large, these policies can have a positive feedback loop. If Social Security contributions are too high relative to benefits, more people will retire as early as possible, driving up contributions even more. 

The robust extension discussed in this section responds much faster because it depends on the rate of change of the forecasting error. As soon as reality diverges from the forecast, this policy responds by raising or lowering rates assuming that the current rate of divergence will continue. If this expectation is true, this robust extension spreads the cost of the forecasting error more evenly over time. However, if the forecasting error oscillates around zero or the divergence slows down, this policy overreacts and unduly penalizes payees early in the policy period. Nevertheless, this extension is less cyclical than the  Pay-As-You-Go solution, because it raises rates as the divergence between forecast and reality accelerates and lowers rates as it slows down. Given that most public policies aim to smooth out the economic cycles, this is more desirable than the  Pay-As-You-Go solution which tends to amplify these cycles.

\section{Concluding Remarks}
\label{sect_summary}

In this paper, we propose a framework to analyze how public policies respond to forecasting errors. We argue that good policy design should minimize sensitivity to forecasting errors, and derive constraints on policies that accomplish this. These constraints show that policies can only be insensitive to forecasting errors if their cashflows depend on parameter and their time derivatives.

Model-independent results, like the constraints derived here, are useful because they can guide policy design. A designer can test a draft policy for robustness using the results in this paper. Conversely, the results can guide a policy designer, by indicating which ingredients must be present for a policy to be insensitive to forecasting errors.

Most current policies are fully exposed to forecasting errors because they do not satisfy the constraints derived here. This may be part of the reason why Social Security, many defined benefit plans, and other entitlements run into financial problems. Policies designed to satisfy the constraints derived here have a feedback mechanism that allows them to adapt if reality unfolds differently than the policy designers anticipated.

The methods developed in this paper should apply in other contexts where parameter-dependent cashflows play a role. Examples may include pension funds, business plans, and similar applications.

\end{document}